\documentclass[preprint,aps,onecolumn]{revtex4}
\usepackage{graphicx}
\usepackage{dcolumn}
\usepackage{bm}
\usepackage{amsfonts}
\usepackage{amsfonts}

\begin{document}
\title{Stability and mode analysis of solar  coronal loops \\ using  
thermodynamic irreversible energy principles II. \\ Modes in twisted non--isothermal magnetic field configurations}
\author{A. Costa }
\email{A. Costa:  acosta@mail.oac.uncor.edu }
\affiliation{Instituto de Astronom\'\i a Te\'orica y Experimental (IATE--CONICET), C\'ordoba, Argentina  } 
\author{ R. Gonz\'alez}
\affiliation{ Universidad
Nacional de General Sarmiento (UNGS), Argentina, Departamento de F\'\i
sica (FCEyN-UBA)  Buenos Aires, Argentina }
\date{}






\begin{abstract}
We study  the stability and the modes of non -- isothermal coronal
loop models with  different intensity values of the equilibrium twisted magnetic field.We use an energy
principle obtained via non -- equilibrium thermodynamic arguments.
The principle is expressed in terms of Hermitian operators and
allows to consider together the coupled system of equations: the
balance of energy equation and the equation of motion, to obtain modes and eigenmodes in a
 spectrum ranging from  short to long--wavelength
disturbances without having to  use weak varying  approximations of
the equilibrium parameters. Long--wavelength perturbations
introduce additional difficulties because the inhomogeneous nature
of the medium determines  disturbances leading to continuous
intervals of eigenfrequencies which cannot be considered as purely 
sinusoidal.We analyze the modification of periods, modes
structure and stability when the helicity, the magnetic field
strength and the radius of the fluxtube are varied.  The efficiency of the damping due to the
resonant absorption mechanism is analyzed in a context of
 modes that can either impulsively release or storage
magnetic energy.We find that the onset of the instability is associated to a critical value of the helicity and that the  magnetic energy content has a  determinant role on 
 the instability of the system with respect to the stabilizing effect of the resonant  absorption mechanism.
\end{abstract}

\maketitle
\section{Introduction}
\subsection{Variational principle}

A crucial
requirement for any theoretical model of coronal structures
is to give account of   the stability and evolution of   far--from--equilibrium states which are responsible of the characteristic rich topology and dynamics of the solar corona.
This implies  to 
consider the coupling of thermal and mechanical  equations.
Different stability analysis of
solar structures can be found in the literature, generally
restricted to special types of perturbations and specific
equilibrium models. These includes, models that consider adiabatic
configuration such as the ones analyzed via the classical
criterion of \citet{ber} or those that
presuppose  static equilibrium and analyze thermal stability. In
the application of Bernstein's
 criterion, the adiabatic assumption implies that the energy balance equation is not required
 and thus dissipation is
impossible. Also the assumption of static models is a strong, and
often unjustified, restriction for open systems.

In this paper we apply an energy
principle to analyze the stability of solar coronal loops when helical modes are present.
The principle was obtained in  previous papers (Paper I:  \citep{cos1};
\citep{cos2}; see also  \citep{us0}) using a general
procedure of irreversible thermodynamics -based on firmly
established thermodynamic laws- that can be understood as an
extension of  Bernstein's MHD principle to situations far from
thermodynamic equilibrium. 

In Paper I and in \citep{cos2}  we showed how to obtain the variational principle for solar
coronal structures
from
the equations that describe the dynamics of the system. The method consists of obtaining a
Lyapunov function,
also known as generalized potential, that represents the
mathematical expression of the stability conditions.  The principle is subject to physically reasonable
requirements of hermiticity and antihermiticity over the matrices.
For a more detailed presentation see Paper I and the references therein.

\subsection{Solar coronal loops}

MHD loop oscillations in the corona are known to be
strongly damped, mostly having  decaying times of few periods
$N_{p}\approx 2 - 7 \  periods.$ While thermal conduction, with the
contribution of
radiative cooling mechanisms, could be the main
cause of the damping of pure
MHD slow magnetoacoustic mode oscillations they are unimportant for the  MHD fast
modes.
Resonant absorption and phase mixing seem more promising
 in giving account of the rapid decay (\citep{gh94}; hereafter HG,
 \citep{go02})  of the ideal fast
 oscillations   of these strongly inhomogeneous and structured plasma systems.
 Inhomogeneous
 equilibrium distributions of plasma density
and temperature varying continuously across the magnetic field led to plasma waves with
continuous intervals of eigenfrequencies. The
occurrence of the Alfv\'en ideal MHD continuum in a thin edge layer
is derived from the highly
anisotropic character of  the fast   magnetoacoustic waves
giving rise to a peak of the amplitudes where the perturbation
develops large gradients and the absorption has maxims. However, there is another type of continuum commonly known as slow magnetosonic continuum associated to the inhomogeneity of the equilibrium parameters along the axis of the loop (see Paper I). This inhomogeneities are associated, for example, to changes in the 
density concentration at the loop basis. If the magnetic field is twisted the inhomogeneities led to the coupling of Alfv\'en and slow magnetosonic continuum modes (\citep{bel}). 

The resonant absorption mechanism of
wave heating consists on the non--dissipative transference  of
wave energy from the collective line-tied wave with fast discrete eigenvalues (kinetic energy of the fast
radial component) to a local resonant  mode in the Alfv\'en continuum, (kinetic energy of the
azimuthal component), which is then dissipated in an enhanced
manner. Then, the  continuum oscillations are converted
into heat by dissipative processes; as the medium has large
gradients in the Alfv\'en speed, the oscillations of neighboring
field lines become out of phase and  shear Alfv\'en waves lead to
enhanced viscous and ohmic dissipation (see \citep{pr83} for the
linear regime and \citep{nak97} for the nonlinear one). The mode
conversion from the collective to the local mode occurs in a  time
that is  non--dissipative and generally much shorter than the
second time scale which is related to the dissipative damping of the
small--scale perturbations of the local mode in the resonance
layer (\citep{rob00}; \citep{van04}). 


The whole temporal pattern description  of
modes that exhibit a combination of global (discrete line--tied
fast eigenmode) and localize (Alfv\'en continuum mode) behaviour
is known as  quasi--mode. Moreover, the mixed nature of the modes
is not only due to the temporal behaviour but also to the
boundary value problem giving rise to a spatial behaviour which is
also of a mixed nature, i.e. coronal loops with line--tying
constraints cannot support pure waves: Alfv\'en, slow or fast
magnetoacoustic modes.  
HG
studied the mixed spectral description of coronal loops (i.e. the
resulting superposition of basic waves which adjust the line--tied
condition) without assuming a straight magnetic field and forcing
the loop to follow the photospheric velocity perturbations. They
found that pure  Alfv\'en and pure slow modes are  obtained as
singular limiting cases of cluster spectra of Alfv\'en--fast or
slow--fast modes, where the fast components are localized in a
photospheric boundary associated to the line--tied condition: 
the coronal part of the
loop acting as a resonant cavity  of large   Alfv\'en
components  and  fast components, with a small but
rapidly varying amplitude, located in the photospheric boundary layer.
They found that heating of coronal loops by resonant absorption is
due to the line--tied Alfv\'en continuum which no  longer depends
on the poloidal magnetic field and that the corresponding
eigenmodes have a global ballooning feature which is characterized
by  an accumulation point given by the Alfv\'en frequency. In \citep{gh93} (hereafter
GH), a variational principle, based in Bernstein's
principle, was obtained to  derive the Alfv\'en and slow continuum
frequencies in a line--tied inhomogeneous cylinder. Stability
considerations led them to conclude the global  stability  of
coronal loops.

In this paper, following results of Paper I  we apply our energy principle to consider the stability and mode structure of loop inhomogeneous coronal models with non--vanishing helicity.  Our principle has the advantages that it does not require a WKB approximation and that, as was mentioned, it  allows the consideration of the coupling of the thermal and mechanical equations that are necessary to analyze far from equilibrium states.

\section{The MHD  stability criterion for coronal structures}

Solar coronal conditions with large Reynolds numbers are well  fitted by
ideal  MHD plasma models  (i.e.  infinite electrical
conductivity $\sigma \gg 1 $ leading to vanishing viscosity and
ohmic dissipation). Thus, the fundamental equations considered are the
mass conservation equation,
the perfect gas law or state equation for a fully
ionized $H$ plasma and the induction equation, with vanishing magnetic diffusivity due to the conductivity properties. The energy balance  equation  takes the form:
\begin{equation}
\frac{\rho^{\gamma}}{(\gamma-1)}\frac{D}{D
t}(\frac{p}{\rho^{\gamma}})=-\nabla\cdot\vec{F_{c}}-L_{r}+H\label{1}
\end{equation}
$\vec{F_{c}}$ is the heat flux due to particle conduction along
the loop, $L_{r}$ is the net radiation flux and $H$ the heating function
which was chosen as in Paper I: $H = h \rho + H_{0}$.
 Eq.~\ref{1} expresses the fact that the gain in particle
energy (internal plus kinetic) is due to the
external heating sources represented by the heating function, heat
flow and radiation losses;
all other heating sources were considered as vanishing terms
implying that the optically thin assumption holds. Note that the non--ideal contribution in the energy equation ($L$) is associated to the open character of the loop system.

Once the linearization  around a nonlinear equilibrium or stationary state is performed, and after a straightforward manipulation procedure where the hermiticity requirements are  fulfilled the generalized energy principle and the respective frequencies are obtained  (Paper I and
\citep{cos2}) as:
\begin{equation}
 \delta^{2} W_{p} =\frac{1}{ 2}\int ( \vec{\xi}^{*} \beta  F \vec{\xi}+T_{1}^{*} AT_{1}
 +T_{1}^{*}B\vec{\xi} -\vec{\xi}^{*}BT_{1})d^{3}x\geq 0.
 \label{2}
\end{equation}
\begin{equation}
\omega^{2}  =- \frac{\int  ( \vec{\xi}^{*}  \beta  F\vec{\xi} +
T_{1}^{*}AT_{1}+T_{1}^{*}B\vec{\xi}-\vec{\xi}^{*}BT_{1}^{*} )
d^{3}x}{\int (\vec{\xi^{*}} \beta \rho_{0}\vec{\xi} )d^{3}x}
\label{3}
\end{equation}
with the same normalization condition as in Paper I. 
$ F$ is the known Bernstein operator for the system, $\xi$ and $T_{1}$ are the motion and temperature perturbations and operators $A$ and $B$ are as in Paper I.
For the non-dissipative cases ($L=0$ or equivalently $T_{1}=0$),  last expressions (discarding the presence of factor $\beta$ which appears in the equations  to fit the  Hermitian and anti--Hermitian conditions) are reduced to the
well--known  Bernstein MHD energy principle and its respective frequencies.  

\section{Application to an   inhomogeneous loop model  with non--vanishing helicity}

On one hand, the azimuthal component of the loop  perturbation is
believed  to be one of the  principle responsible of resonant
absorption and damping of  ideal oscillations; on
the other, this component is associated to the storage of
magnetic energy in systems with non--vanishing helicity which
eventually is released by instabilities. Thus, we are interested in
analyzing the changes produced in the stability of
non--homogeneous loops subject to helical perturbations. This is, loops
with inhomogeneous distributions of plasma density and
temperatures subject to body modes and with non--vanishing  helicity. In this case, the
Alfv\'en, slow and fast magnetoacoustic  cylinder
modes cannot longer be associated to the azimuthal,  longitudinal
and radial  components respectively.  The
observational importance of helical modes cannot be neglected
and it is poorly known how helicity affects important physical
features of  mode oscillations (e.g., damping mechanisms,
stability and periods). However, a mode classification can be accomplished
via the analysis of the  mode variations, described in an orthogonal basis, while helicity is varied. The basis  is formed by  the orthogonal displacements:
parallel and perpendicular to the magnetic field  and  the radial
(and perpendicular to the surface of the tube) one, of
observational interest.

The fundamental modes are generally observationally and
energetically more important than their harmonics. For these
global modes the inhomogeneous nature of the medium cannot be
ignored and it determines the structure  of the disturbance which
cannot be taken as sinusoidal, making the traditional normal mode
analysis useless for this treatment (sinusoidal dependence with constant coefficients), i.e. at least a WKB approximation, of weakly varying parameters compared to a typical wavelength, is required. Moreover, the occurrence of
either an infinitely degenerate eigenvalue or an accumulation
point giving  rise to a continuous spectrum are associated to inhomogeneities. 
We consider two types of inhomogeneities: the inhomogeneity of the equilibrium parameters along the loop axis, and the inhomogeneity across the loop axis when the radius is varied.  As a first order approximation we neglect the effect
of gravitational stratification and thus confine  the analysis to
characteristic spatial scales lower than the pressure scale height
in the solar corona. In order to analyze the stability and to
obtain the frequencies and modes the physical quantities in
eq.~\ref{2} and eq.~\ref{3} must be calculated along the loop
structure.

\subsection{Mechanical equilibrium}

To determine an equilibrium configuration we assume
force--free equations. This assumption is  justified for coronal conditions due to the fact that in plasmas with low
$\mathcal{\beta}$ (gas pressure over the magnetic pressure) the
pressure gradient can be neglected in comparison to the Lorentz
force. For the chromosphere and the photosphere the force--free approximation  may not be a good one. However, it is a widespread supposition \citep{rud}:
  perturbed systems are believed to relax to  new force-free, minimum energy states  and  chromospheric conditions seem to be well fitted to force--free models from $4. \ 10^{5} \ m$ \citep{asc4} (Chapter 5).

Coronal loops are generally modeled as thin cylindrical
fluxtubes where the curvature and related forces can be neglected so
the cylindrical geometry can be applied. The fluxtube is assumed
as line--tied to the photospheric plasma through its footpoints
which are forced to follow the photospheric velocity
perturbations. The random velocity field creates vorticity
generally twisting the coronal fluxtubes. Thus, a relation between
the helical twist and the force--free parameter can be derived as
follows (e.g. \citep{stu94}). The coronal loop model is obtained
from the equations

\begin{equation}
\nabla \times \mathbf{B_{0}}=\alpha(r) \mathbf{B_{0}} \ \ \ \mathbf{j} \times \mathbf{B_{0}}=0. \label{3}
\end{equation}
Also, since $\mathbf{B_{0}}$ is force--free, $\nabla p=0$ everywhere and thus  has
a constant value along the loop. We consider a straight cylinder with a nonuniform distribution of density and temperature and a resulting uniform
twist over an initially non--rotated field $ \textbf{B}=(0,0,B_{z})$ yielding
 the unperturbed
magnetic field  

$$\mathbf{{B_{0}}}= (B_{r},B_{\phi},B_{z})=B_{0}(0, \frac{br}{\Delta}\frac{1}{\Delta})$$

\noindent with $\Delta=1+b^{2}r^{2}$  and $ b=2 \Pi N_{t}/L$ ($
N_{t}$ number of turns over the cylinder length $ L$). Then,
\begin{equation}
\frac{B_{\phi}}{B_{z}}=\frac{r \partial \phi}{\partial z}=\frac{r 2 \pi  N_{t}}{L}=br \\
\alpha(r) =\frac{2b}{\Delta}\label{4}
\end{equation}
We assume a given value of
the cylinder radius  $r=R$, thus the line element results  a function of
the coordinate
$z$: $s=s(z)$. The dependence with the radial component will be taken into account by  considering different values of the radius $R$.

\begin{equation}
ds^{2}=R^{2}d \phi^{2}+dz^{2}=\left( 1+R^{2} b^{2}\right) dz^{2} =\Delta dz^{2}\label{5}
\end{equation}

\subsection{Thermal equilibrium}

The thermal equilibrium is obtained, as in Paper I, assuming  $L=0$ in the balance energy equation 
(eq.~\ref{1}) . The procedure
developed consists in obtaining the function of the temperature along the
arc element $s$ by integrating eq.~\ref{1} with the constraint $L=0$ and
 replacing border conditions: the temperature at the bottom $T_{b}=10^{4}K$ and the temperature at the top $T_{t}=10^{6}K$. The known expression (see chapter 6 of
\citet{pri}) is obtained
\begin{equation}
\left[\frac{dT}{ds}\right]^{2}=\frac{p^{2}\chi}{2k_{B}^{2}k_{0}(\alpha
+\frac{3}{2})}T^{\alpha-\frac{7}{2}} \left[1 - (
\frac{T}{T_{t}})^{2-\alpha}\right]\label{6}
\end{equation}
which has to be inverted to obtain
$T=f^{-1}(s)$ \citep{ar95} as
\begin{equation}
\frac{dT}{ds}=\mathcal{A}\left[\frac{d\mathbb{B}_{v}}{dv}\frac{dv}{dT}\right]^{-1}    \ \ \  where  \ \ \ \mathbb{B}_{v}(\frac{1}{2},q)=\int_{0}^{v}t^{p-1}(1-t)^{q-1}dt  \label{7}
\end{equation}

\noindent 
 with

$$p=  \frac{1}{2};    v=1-(\frac{T}{T_{t}})^{2-\alpha};  q=(\frac{\alpha}{2}+\frac{3}{4}) (2-\alpha)+1  $$

$$ \mathcal{A}=(2-\alpha)T_{t}^{\frac{\alpha}{2}-\frac{11}{4}}((p^{2}\chi)/(2k_{0}
(\alpha+\frac{3}{2})k_{B}^{2}))^{\frac{1}{2}}.$$
We use $\alpha=-\frac{1}{2}$ so $q=\frac{6}{5}$
to numerically calculate 
 the modes, \\
$ s=\frac{1}{\mathcal{A}}\mathbb{B}_{v}(\frac{1}{2}, \frac{6} {5})\rightarrow\mathcal{A}=\frac{5}{2}T_{t}^{3}(\frac{p^{2}\chi} {2k_{0}k_{B}^{2}})^{1/2}.$\\
Also,  from boundary conditions $\upsilon =0$, thus  the constant value of the  heating function results 
$H_{0}= 7
p^{2} \chi T^{\alpha -2}_{t}/\left(8 k_{B}^{2}(\alpha
+\frac{3}{2})\right).$ 

\subsection{The perturbation}

To calculate the stability and the structure of the  modes the
general perturbation  along the equilibrium magnetic field  is
written
\begin{equation}
\vec{\xi}=[\zeta_{r}(r,z)
\mathbf{e}_{t}+i \zeta_{\phi}(r,z) \mathbf{e}_{\phi}+\zeta_{z}(r,z) \mathbf{e}_{z}]e^{im\phi} \ \ T_{1}=T_{1}(r,z)e^{im\phi}\label{9}
\end{equation}
with $r=R$.
The  $\phi$ dependence only   appears  in the exponents that multiply the perturbation; the integration with respect to this coordinate is straightforward.
Then, representing the equilibrium functions of the different
quantities with a 0 sub-index,  defining 

\noindent
$$ \mathbf{e}_{t}=(Rb \mathbf{e}_{\phi}+ \mathbf{e}_{z} )/ \sqrt{\Delta} \ \ \  \nabla_{\parallel}=\mathbf{e}_{t}(\mathbf{e}_{t}\cdot\nabla)\nonumber \ \ \  \rho_{t}=\frac{m_{p}}{k_{B}T_{t}}$$

\noindent
 with $\mathbf{e}_{\phi}, \mathbf{e}_{z}$ the cylindrical versors and $\mathbf{e}_{t}$ the tangential versor, we obtain   a non--dimensional expression for the energy principle
 of eq.~\ref{2}:
\begin{equation}
\delta^{2} W_{p}= \delta^{2} W_{c}+\delta^{2} W_{m}+\delta^{2} W_{hc}+\delta^{2} W_{r} 
\label{10}
\end{equation}
where $\delta^{2} W_{c}$ is the generalized potential energy associated to compressional terms, $\delta^{2} W_{m}$ corresponds to the magnetic contributions,  $\delta^{2} W_{hc}$ corresponds to the heat conduction terms and $\delta^{2} W_{r}$ to the radiative contributions. The explicit form of these functions are given in the Appendix. The Bernstein's generalized potential energy corresponds to the magnetic contribution and part of the compressional one. In the generalized version of the energy principle  additional terms appear in the $\delta^{2} W_{c}$ term and also $\delta^{2} W_{hc}$ and $\delta^{2} W_{r}$ are entirely new terms.

\section{Results and discussion}

Convective motion of the photosphere  is believed to provide the
energy that is storage in twisted magnetic coronal fields allowing
the presence of long--lived coronal structures until it is
released by instabilities  (\citep{raa};  \citep{vrs}). On the
other hand, continuous spectra are generally associated to stability. 
An accepted conjecture  establishes that unstable
modes have  a discrete spectrum (see
 \citep{fre} or  \citep{pri}). There are two types of  possible continuous spectra in this problem. The inhomogeneous character of the equilibrium parameters along the loop axis can lead to a continuum that couples to the Alfv\'en continuum
\citep{bel}; e.g, when the disturbances considered are comparable to the inhomogeneous characteristic wavelength stable eigenvalues    can give rise to a  continuous
spectrum  ($L/2$, the equilibrium structure in the $z$ component). This is the case studied in Paper I. On the other hand, GH
established, for non vanishing helicity systems, that 
there is a continuous spectrum associated with the line-tied Alfv\'en
resonance leading to the  damping and heating by the resonant absorption mechanism and thus,  directly  relate to
the stability of loops.  They  also pointed out how to  obtain  the
resonant singular limit $\omega$, from the class of physically
permissible solutions,
\begin{equation}
\omega(r)=\frac{nB_{z}(r)}{\int_{-L}^{L}\sqrt{\rho(z)}dz}.
\label{11}
\end{equation}
 This resonance results because of the absence
  of an explicit dependence on the azimuthal magnetic field component ($B_{\varphi}$).

Thus, in order  to understand in which conditions which  mechanism can dominate
and give account of the different 
 scenarios i.e., the driving of the
 instability or the damping of mode oscillations, it is critical to gain
 knowledge about the dynamics and energetic contribution of twisted  structures.
Yet, the implications of the twisting in  theoretical and
observational descriptions are poorly known; e.g., there is no
clarity about  the modification of the dispersion relation  and observational data
are   indirectly inferred.

In this paper we focused our attention to describe the changes in
periods, stability and mode structure  of coronal loops when the
helicity, the magnetic field intensity and the radius are varied. For loops
with vanishing  helicity it is well established that  the Alfv\'en
line--tied resonance continuum is responsible of the damping of
kink ($m=1$) quasi--modes via the transfer of energy from the
radial component into the azimuthal one, i.e., from discrete
global modes into the  local continuum modes where phase mixing
can take place. Still, the twisting of the magnetic field leads to
the coupling of MHD cylindrical modes making difficult to provide
a classification in terms of the behavior of pure--like modes.

In order to calculate modes and frequencies we followed the
schematic procedure described in Paper I and in
\citep{gal1}. We used a symbolic manipulation program to integrate
the equations. $\delta^{2}W_{p}$ and the perturbations were
expanded in a six dimensional--Fourier basis on the independent
 coordinate $z$ that adjusts to
border conditions, i.e., the four perturbated components  (eq.~\ref{9})
were expanded in a six mode basis to
 obtain $24$ eigenvalues and eigenvectors for each of the  helicity
  and the magnetic field values. 
Only  the first
   eighteen eigenvalues were considered (the others are more than two order of magnitude smaller
     and accumulate at zero; the eigenvectors are also
     vanishingly small). 
Thus, a quadratic form for $\delta^{2}W_{p}$
was obtained and was minimized with the Ritz variational
procedure. A matrix discrete eigenvalue problem subject to a
normalization constraint was obtained. From the resulting modes and   the
generalized potential energy (eq.~\ref{2}): $\delta^{2}W_{p}\geq0$ the stability of each mode was determined.

The coronal loop parameters used were: $L=10^{10}cm$ (or $L=100Mm$),
$T_{b}=10^{4}K$ $T_{t}=10^{6}K$ $n_{e}=10^{8}cm^{-3}$ electron
number density $p_{t} =2k_{B}T_{t} \;$;
$\rho_{t}=m_{p}p_{t}/k_{B}T_{t}$. 
Frequencies and modes were calculated for
two different values of the magnetic field: $B_{0}=10G$ and
$B_{0}=100G$, and for three different values of the  helicity $b=(3.1 \ 10^{-8}  ; \ 3.1
 \ 10^{-7}; \ 1.9  \ 10^{-6})$  which correspond to the adimentional values:
 $b_{a}=(2.8; \ 28; \  170)$  with $N_{t}=(0.45;  \ 4.3;  \ 13.7)$, $N_{t}:$ the
  number of turns
over the cylinder length. These helicity values defined as  weak,
moderated and strong helicity respectively correspond to the
classification given in  \citep{asc4} (Chapter 5).
 The adimentional radius
was initially chosen as $R=0.01$.
In what
follows we summarize the conclusions obtained from the data analysis which are displayed in  three
tables.


Table 1 shows the periods (in minutes) for weak, moderate and strong helicity
 for two values of the magnetic field intensity ($B_{0}=10G$ and  $B_{0}=100G$
(left and right panel respectively).  S and U
letters indicate the stable--unstable character of the modes.
From the table we see that:

\noindent
I) Weak helicity modes are stable. This  is in accordance with the analytic results by  \citet{rud} who studied nonaxisymmetric oscillations of a thin twisted magnetic tube with fixed ends in a zero-beta plasma.

\noindent
II) Higher modes have an accumulation point at zero, indicating  the presence of
    a continuum  spectra of stable modes (as in Paper I).  Note that,  calculus performed
 via discrete basis, as in our case, give spectra that are necessarily discrete.
  Thus, an
accumulation of discrete eigenvalues 
suggests a stable continuum  spectrum. 


\noindent
III) The  $B_{0}=10G$ case has larger periods than the $B_{0}=100G$ one.
    For moderate and strong helicity the eigenvalues of the first panel  follow a scaling
  law with that of the second one i.e., they scale with the magnetic field intensity exactly as  the Alfv\'en speed does $P_{10G}\simeq 10 P_{100G}$. 

\noindent
IV) As HG and GH, we note a clustering of the spectra associated to the
  change from real to imaginary eigenvalues (and viceversa). There is a pronounced change (in the spacing of the periods or/and in the stability) from the sixth mode  to the seventh mode. This is noted by a double line in Table 1 and related to the importance of the parallel  component with respect to the perpendicular component (see Table 2). 
Up to period number ten real -- imaginary eigenvalues of the first panel
 ($B_{0}=10G$) correspond to real--imaginary  ones of the second panel ($B_{0}=100G$). Also,
excepting large order periods $n>10$,  when
      the helicity is increased from weak to moderate  the imaginary stable
       eigenvalues turn to imaginary unstable ones. For $B_{0}=10G$  and weak
   and moderate helicity cases there are five different groups of periods
   ($P_{1}-P_{6};P_{7}-P_{10};P_{11}-P_{12};P_{13}-P_{14};P_{15}-
P_{18}$) (see also Table 2). The clustering is more difficult to establish i.e., the
differences are less pronounced, with increasing magnetic field
intensity and larger order periods.

\noindent
 In order to compare our results with those given by these authors
 we calculated  the expression eq.~\ref{11} for our modes. We found that, all periods  excepting  $P_{1}-P_{6}$   weak helicity modes satisfy 
  the relation and thus, they belong to the Alfv\'en continuum spectrum
justifying the scaling law described in III.
   As HG, we conclude that the change in the real--complex character of  the
    $P_{6}-P_{7}$ eigenvalues is associated to the existence of an accumulation
     point of the resonant Alfv\'en continuum, however we find that this change is not necessarily related to a change 
      in the stability as they claimed.  Note that all modes  with weak
      helicity are stable (even the imaginary ones); in all the other cases the imaginary character of the eigenvalues is associated to instability. 
Yet, the continuum stable eigenvalue conjecture
        is here still valid  \citep{fre},  \citep{pri}; it applies to a spectrum with an accumulation point in zero; we found stable modes for all the helicity values and for the two magnetic field values with $P_{n>14}$. Note that the analysis of stable modes  is still of interest because depending on the relative characteristic times of stable and unstable modes  the stable ones could   be active 
 and accessible to observations.

\noindent
The presence of at least one  unstable mode means that the equilibrium
 state is unstable. Thus, taking into account the whole range of stable modes, we confirm previous results
leading to conclude that field configurations with some
degree of twisting  give a stabilizing effect allowing the storage
of magnetic energy  \citep{raa2}, i.e.,   when the helicity is augmented the stable weak case  turns to an unstable one suggesting a critical value. 

\bigskip

In Paper I,  we obtained  only one unstable mode classified as  slow
magnetoacoustic mode due to the almost longitudinal character (parallel to the
 magnetic field) of the wavevector  perturbation
  and to the fact that the period did not changed with the
  intensity of the magnetic field,  resembling  acoustic waves with sound speeds,
    $v_{s}$,  independent of the magnetic field. The characteristic
     unstable time obtained in Paper I was $\tau_{u}=36 \ min$, corresponding
      to a typical slow magnetoacoustic fundamental  period with a characteristic
       wavelength  of the order of the loop length $L/2$.  Also,
       we obtained a continuous set of stable modes classified as
        fast magnetoacoustic modes due to their large value  component orthogonal
          to the magnetic field and to the fact that the eigenvalues scale with
           the intensity of
 the magnetic field  as   $P_{11G}\simeq 10 P_{100G}$; thus resembling
 the dependence of the
    Alfv\'en waves $v_{A} \sim B_{0}$.

Table 2 (First Panel) displays the resulting features associated to the
relative intensity of the parallel and perpendicular to the field components
($(\xi_{\parallel},\xi_{\perp},\xi_{r}) $ is an orthogonal basis)
and their classification  as  slow--like (S) or  fast--like (F). The relative
phase between the  components is also indicated in the table
by P (in phase) and IP (inverted phase). Table 2 (Second Panel) also shows the
intensity relationship between the cylindrical components. 
In order to classify the modes and to compare 
 with  the slow and fast
magnetoacoustic modes obtained in Paper I, we calculated  the cylindrical
 mode components and also the tangential and normal to the field
  components
   ($\xi_{\parallel}=
(Rb\xi_{\phi}+\xi_{z})/\Delta; \xi_{\perp}=
(\xi_{\phi}-Rb\xi_{z})/\Delta$).
   Our interest in the $\xi_{\parallel}$,
    $\xi_{z}$, $\xi_{r}$, $\xi_{\perp}$
     and $\xi_{\phi}$ components  resides in that: First,  when  the helicity is weak, the $\xi_{\parallel}$
       component is expected to play the slow-mode role of $\xi_{z}$
        in Paper I.
          Second, the $\xi_{r}$ component is related to the
           fast modes  and determines
            the resonant absorption mechanism when uniform cylindrical
             flux tubes are considered by
              the transferring of energy 
              to the $\xi_{\phi}$  component. When helicity and inhomogeneous distribution of equilibrium parameters  are present it is worth investigating the transferring
               of energy from the $\xi_{r}$ component to the others. In this
                case the resonant damping of global oscillations
                will occur by conversion of kinetic energy of the
                 radial component into kinetic energy of the
                  $\xi_{\parallel}$ and   $\xi_{\perp}$ components;
                  both components forming the plane orthogonal to
                   $\xi_{r}$, and equal to the plane formed by $\xi_{\phi}$ and $\xi_{z}$.

From the  analysis of the amplitude  of the components of the $P_{1}-P_{6}$ modes with respect to the $P_{7}-P_{18}$ ones in the weak helicity case i.e., real and imaginary eigenvector respectively, we could classify the first ones as slow-like modes because: I) their tangential components $\xi_{\parallel}$
are at least an order of magnitude larger
    than the normal ones $\xi_{\perp}$; II) as the helicity is
     weak $\xi_{\parallel}\approx \xi_{z}$ and 
      $\xi_{r}\rightarrow 0; \xi_{\phi}\rightarrow 0$, the
      wavevector is almost tangential to the magnetic field; III) they have 
a larger characteristic time and a shorter characteristic speed than the imaginary eigenvectors. On the contrary, imaginary eigenvalues  are associated to  large values of
the $\xi_{r}$ component and $\xi_{\perp}$  component (due to large values of   $\xi_{\phi}$ (see Table 2 Second Panel)) , and small values of
the $\xi_{\parallel}$ and $\xi_{z}$ components.
 As in Paper I,  when the eigenvalues change form real to imaginary
  the period strongly diminishes and  a change in
  the type of mode  from the slow  to fast magnetoacoustic type occurs. In opposition to Paper I where the
acoustic mode has the same eigenvalue for both magnetic field
intensities, here the modes are affected by the strengthening of
the magnetic field leading to an-order-of-magnitude shorter  period than in the non--helicity case. 
The $\xi_{\parallel}$ and $\xi_{\perp}$ components are in an inverted phase for real eigenvector modes and in phase for imaginary eigenvector modes.

For moderate helicity the overall description is similar but  all the cases having  non vanishing
 $\xi_{\phi}$ component and  all the periods in the resonant
 line-tied continuum. As was mentioned, real--imaginary eigenvalues
 correspond to stable--unstable behavior.

In the strong helicity case, as the weak and moderate ones, we
note for $P_{1}-P_{6}$ larger, but comparable,  values of  the
$\xi_{\parallel}$ component with respect to the $\xi_{\perp}$
component. In this case  the two components of the mode are in phase.
This relationship between the $\xi_{\parallel}$ and $\xi_{\perp}$
components of Table 2 (FP), and their associated phases is found again in the
 modes with $P_{15} - P_{18}$. In spite that  these features are associated to the slow
magnetoacoustic characterization, Table 2 (SP) shows that as $\xi_{z}$ is
vanishingly small, the strong helicity case cannot be classify as a
slow mode. 

\bigskip

When helicity is present the mixed character of the modes manifests  itself making difficult to identify the components that are involved in   the damping mechanism. However, taking into account  the resonant frequency of     eq.~\ref{11}, we noted that (HG)  all  the modes, except those  with  $P_{1}-P_{6}$
 periods of the weak helicity case, have  resonant frequencies suggesting
         that resonant
         absorption in helical modes is associated to modes with significant values of
         $\xi_{\perp}$ component. If this argument is correct we can affirm
          that  the damping mechanism of body helical modes
              is associated to the transfer of kinetic energy of the radial component into
            kinetic
             energy of the $\xi_{\perp}$ component which is not only related to the $\xi_{\phi}$
              cylindrical contribution but also to the $\xi_{z}$ one
                          by the expression
                          $ \xi_{\perp}=(\xi_{\phi}-Rb\xi_{z}) \Delta$.

\bigskip

We also analyzed the change of the period as a function of the radius for different values of the helicity. 
  We found that, for weak helicity,  the increasing of the radius leads to a decrease of the periods. This is in accordance with observations, e.g., observed sausage modes are associated with thicker and denser loop structures and lower periods; while in  other case (unstable cases) the increasing of the radius leads to an increase of the period. 

\bigskip

Table 3 -First and Second Panel- shows the variation of the radius $R$ with the twist
$bR$ for weak and moderate helicity respectively. 
 \citep{rud} has conjectured that the line-tying condition at the tube ends should stabilize the tube and has suggested a critical value ($\sim L b<q$; with $q$ a positive constant and $L$ the loop length) for the onset of instability. Also,
  \citep{lin}  found that  when the  helicity grows beyond a critical value, the kink isolated twisted magnetic flux tubes below the photosphere become unstable. In fact,  Table 3 can be seen as  the variation of $R$ with  the  twist value: $bR$, for     constant values of the helicity $b$ in the two cases:  weak and moderate  respectively. Stability is guaranteed when the loop radius is varied  between $R=0.01$ and $R=0.1$ and the helicity is weak $b=0.05$ (for almost the same value of the length of the loop, $L$). However, when the helicity is incremented to
 $b=0.5$ even for the radius of $R=0.01$ the loop structure is unstable, thus, instability can be associated with the presence of helicity values larger than  a critical one. 

 \bigskip

\begin{figure*}
   \includegraphics[width=4.3cm]{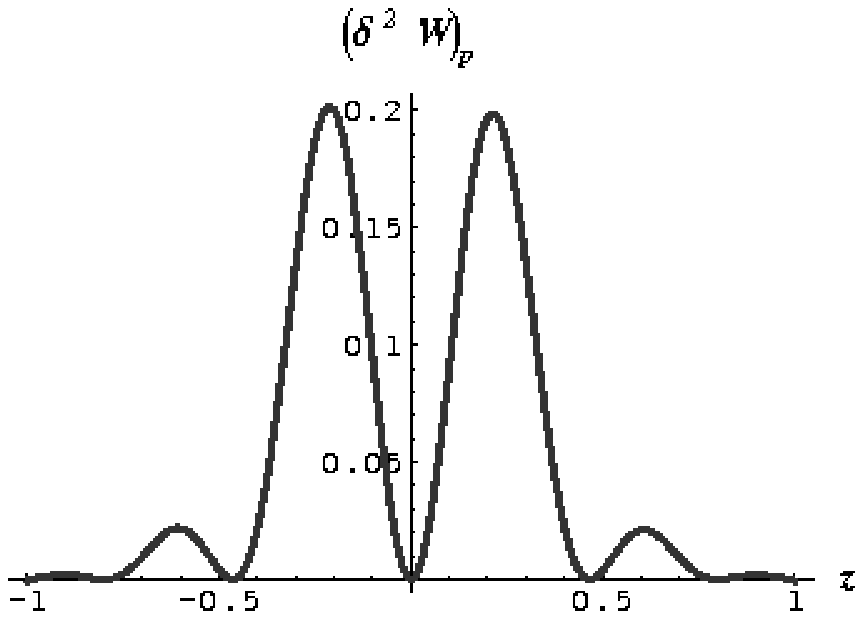}
    \includegraphics[width=4.3cm]{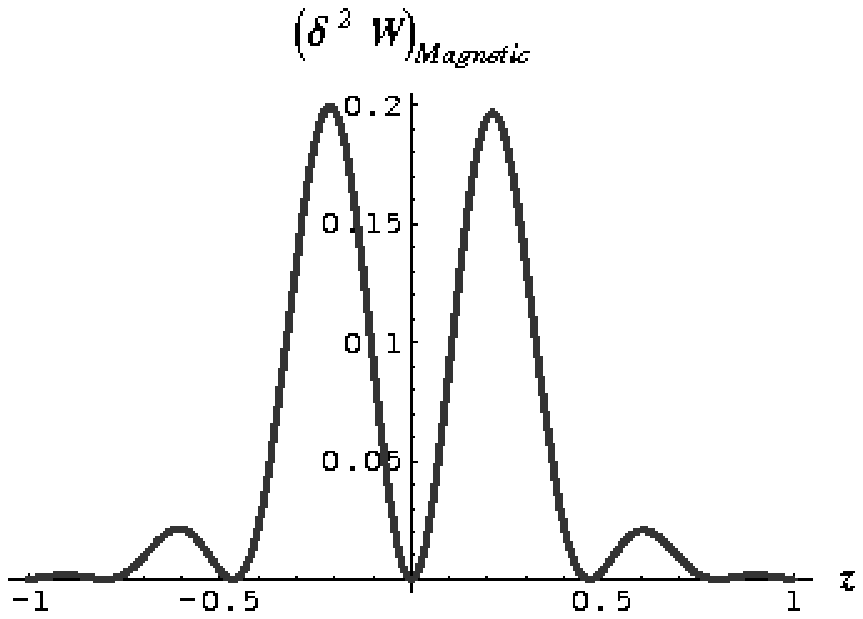}
   \includegraphics[width=4.3cm]{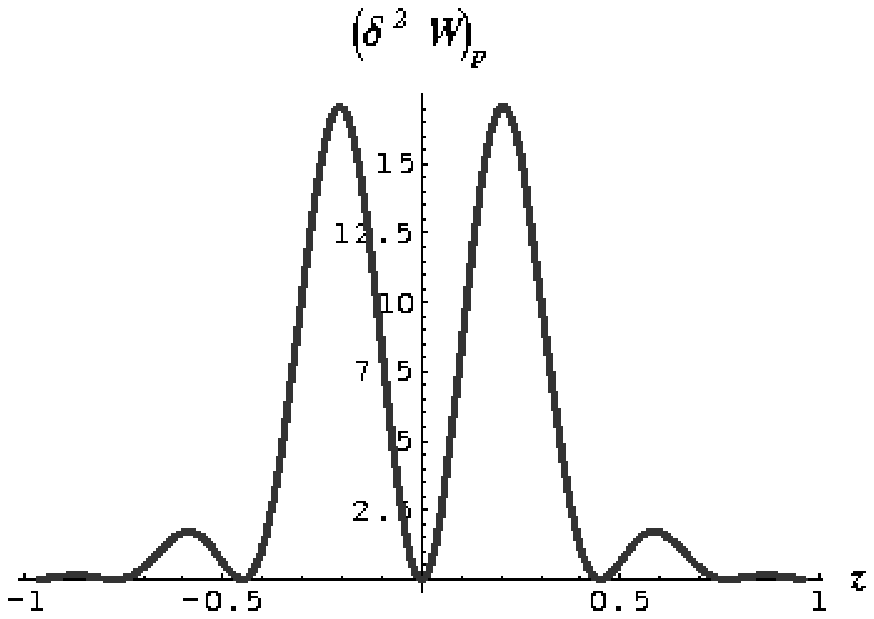}
    \includegraphics[width=4.3cm]{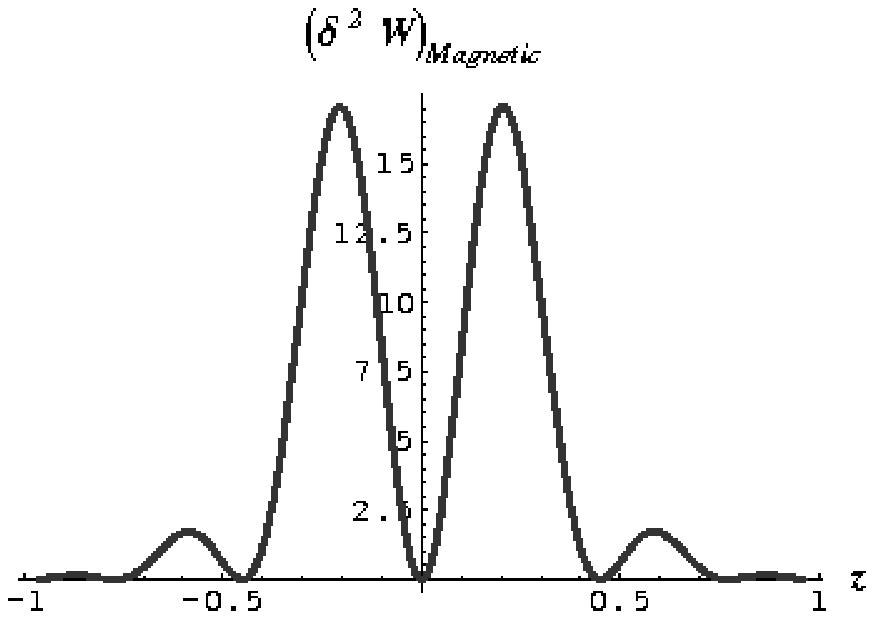}
   \includegraphics[width=4.3cm]{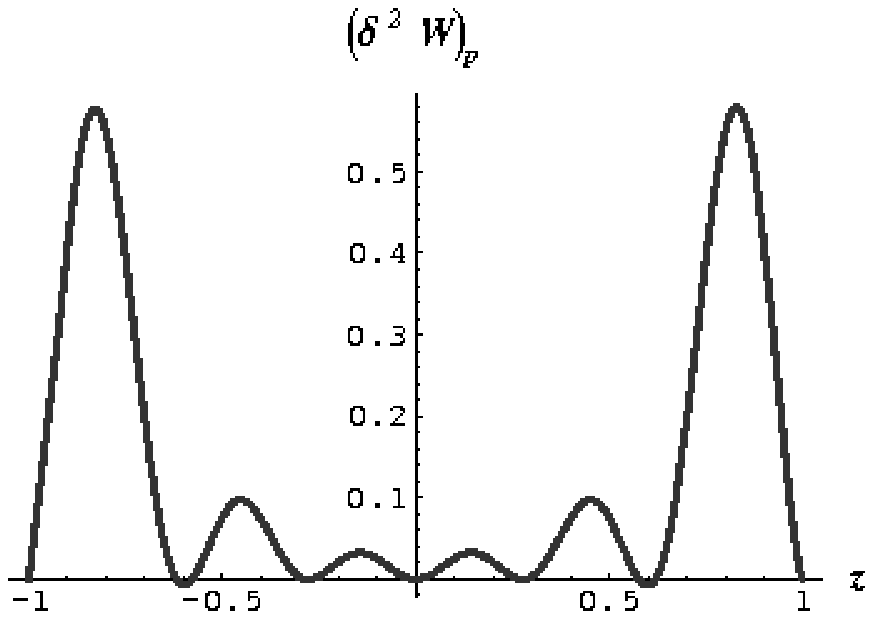}
    \includegraphics[width=4.3cm]{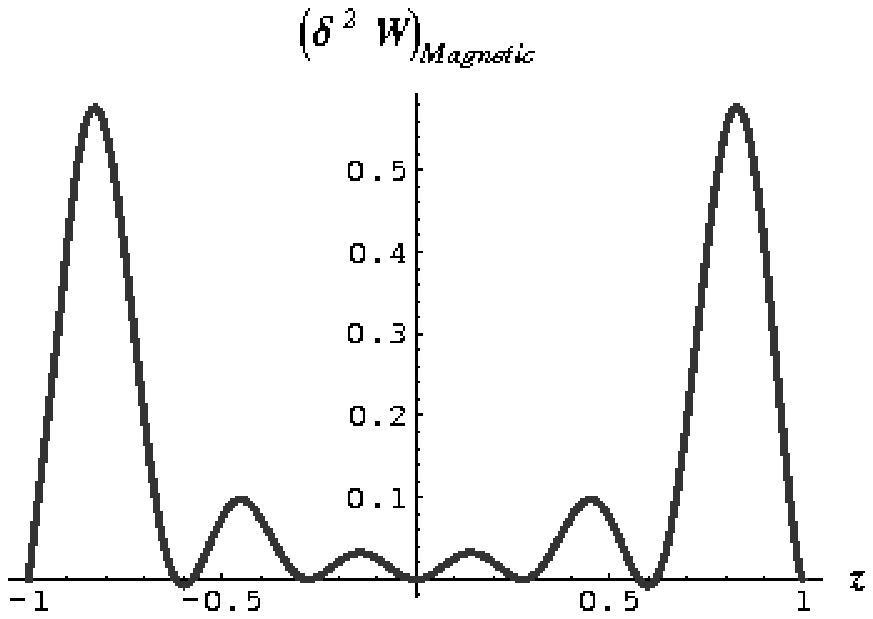}
   \includegraphics[width=4.3cm]{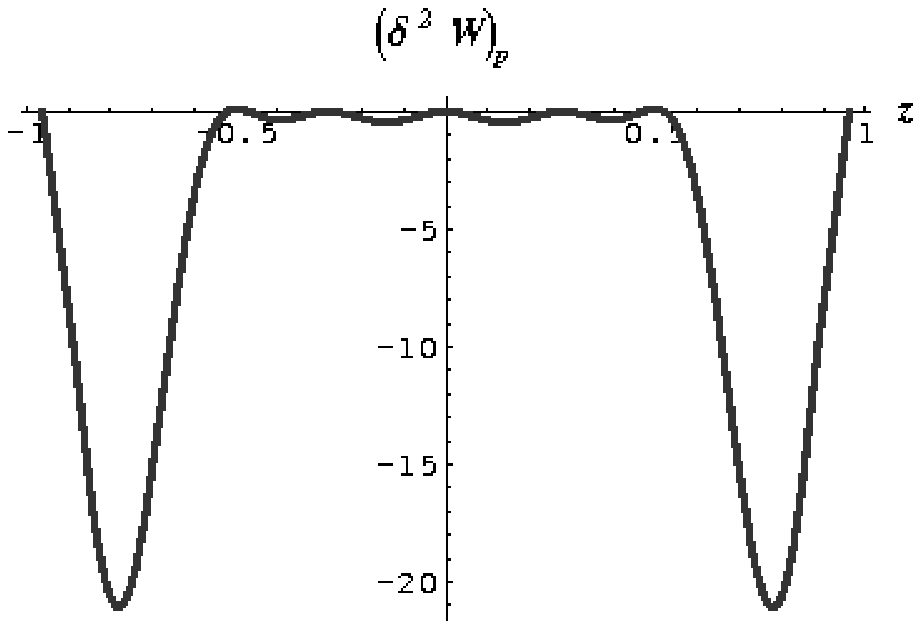}
    \includegraphics[width=4.3cm]{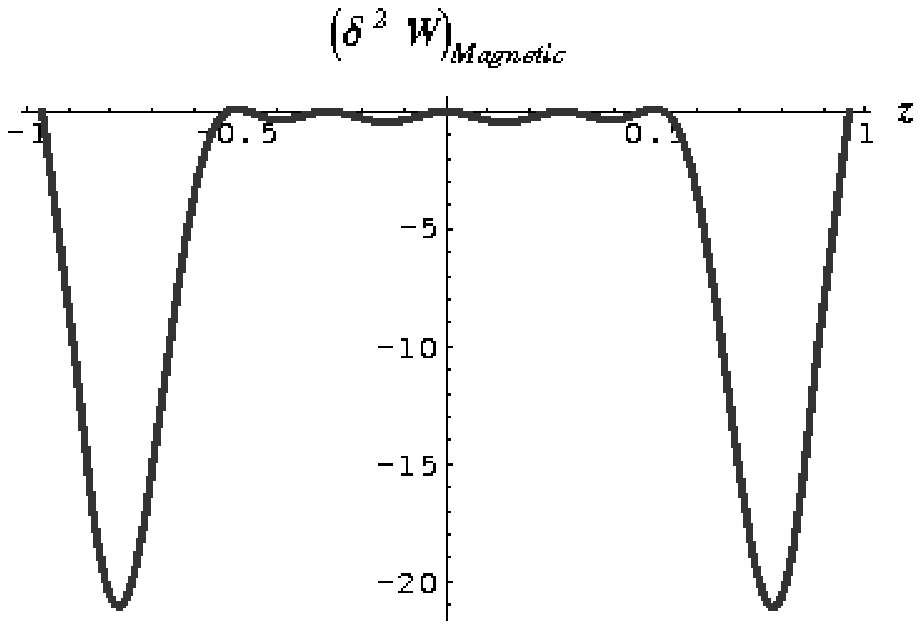}
       \caption{Energy content of the sixth and seventh mode for $B_{0}=10G$. a) Total potential energy and b) magnetic potential energy respectively for the sixth mode $P_{6}=1.23 \ min$ and for weak helicity. c) Total potential energy and d) magnetic potential energy respectively for the sixth mode $P_{6}=1.23 \ min$ and for moderate helicity. 
  e) Total potential energy and f) magnetic potential energy respectively for the seventh mode $P_{7}=0.07 \ min$ and for weak helicity. g) Total potential energy and h) magnetic potential energy respectively for the sixth mode $P_{7}=0.07 \ min$ and for moderate helicity.}
   \label{fig:tres}
   \end{figure*}

 Figure~\ref{fig:tres} shows the general potential energy for  $P_{6}$ and
$P_{7}$ in the weak and moderate cases. Note the change
of this function when the system turns from stable to unstable, as
helicity is augmented i.e., from  $\delta^{2} W_{p}>0$ to
$\delta^{2} W_{p}<0$. Figure~\ref{fig:tres}a and  Figure~\ref{fig:tres}c
display    the total energy composed by the compressional,
radiative, thermal and magnetic energy contributions of   $P_{6}$ mode in the weak and moderate case respectively. 
The same features but for the $P_{7}$ mode  are shown in  Figure~\ref{fig:tres}e and  Figure~\ref{fig:tres}g.  Figure~\ref{fig:tres}b and  Figure~\ref{fig:tres}d show the
magnetic energy content alone for $P_{6}$ mode in the weak and moderate case respectively.  Figure~\ref{fig:tres}f and  Figure~\ref{fig:tres}h show the magnetic energy content for $P_{7}$ mode and for the weak and moderate case respectively.   It can be seen, in this and in all
the other cases, that the magnetic energy content has a  determinant role on 
 the stability--instability of the system, i.e., the stability changes when the magnetic generalized potential energy changes sign. Thus, a result of
this analysis is that the stability of twisted coronal loops is
fundamentally determined by the storing of magnetic energy,  being the
other contributions less significant. Meanwhile, when the helicity is weak or vanishingly small and the magnetic contribution has a stabilizing effect  the other non-dominant contributions, as the non-adiabatic ones,  can play an important  role. This makes possible, for example, the damping of fast excitations due to resonant absorption. Yet, even when one of these contributions is unstable, stable modes could be active for a while if their characteristic periods  are shorter than the characteristic time  of the instability. This is the case of Paper I, where we obtained a slow   mode with an unstable characteristic times of $\tau\sim 36 \ min$ coexisting with  stable fast modes with periods about $P\sim 1 \ min$; moreover, we showed that the instability can be nonlinearly saturated giving rise to a limit-cycle solutions, i.e., an oscillation between
 parallel plasma kinetic energy and plasma internal energy where
 the magnetic energy plays no relevant  role.  
Thus, the
contribution to the stable--unstable character of the modes is
mostly due to the magnetic energy content and not to other
energetic contributions. 
Note that as the balance energy equation  takes into account non--adiabatic contributions, i.e., radiation, heat flow and heat function (with $L=0$ at the equilibrium), the resulting perturbations are not constrained to  the force--free condition. So, one result of the analysis is that the pertubation energy contribution is mainly due to magnetic forces. 
Thus, for these type of twisted magnetic field
models, non--adiabatic perturbations (e.g. thermal perturbations) and resonant absorption seem unimportant to guarantee stability; a loop system with weak 
 storage of magnetic energy (low values of the
helicity)  could be  released if the helicity is suddenly
increased, e.g., by footpoint motions. Meanwhile all the "zoo" of the coronal seismology can be active and accessible to observations.

\begin{table*}
\begin{tabular}{cccccccc}
\hline
$P_{i}$&$weak $&$moderate$&$strong$&$ \ \ \ $& $weak$&$moderate$&$strong$ \\  \hline
$P_{1}$&$1.921 \ S $&$0.209 \ S$&$0.525 \ i \ U$&$ \ \ \ $&$0.159 \ S$&$0.021 \ S$&$0.052 \ i \ U$\\  \hline
$P_{2}$&$1.869 \ S $&$0.204 \ S$&$0.450 \ S$&$ \ \ \  $& $0.158 \ S$&$0.020 \ S$&$0.044 \ S$\\  \hline
$P_{3}$&$1.535 \ S $&$0.169 \ S$&$0.430 \ S$&$ \ \ \  $& $0.154 \ S$&$0.017 \ S$& $0.042 \ S$ \\  \hline
$P_{4}$&$1.533 \ S $&$0.168 \ S$&$0.424 \ i \ U$&$ \ \ \  $&$0.153 \ S$&$0.0167 \ S$&$0.042 \ i \ U$\\  \hline
$P_{5}$&$1.306 \ S  $&$0.143 \ S$&$0.206 \ i \ U$& $ \ \ \  $&$0.151 \ S$&$0.014 \ S$&$0.020 \ i \ U$\\  \hline
$P_{6}$&$1.228 \ S  $&$0.135 \ S$&$0.177  \ i \ U$&$ \ \ \  $& $0.15 \ S $&$0.013 \ S$&$0.017 \ i \ U$\\  \hline \hline
$P_{7}$&$0.068  \ i \ S$&$0.070 \ i \ U$&$0.125 \ S$&$ \ \ \  $& $0.0047 \ i \ S$&$0.007 \ i \ U$&$0.0125 \ S$\\  \hline
$P_{8}$&$0.064  \ i \ S$&$0.066 \ i  \ U$&$0.122 \ S$&$ \ \ \  $& $0.0046 \ i \ S$&$0.006 \ i  \ U$&$0.012 \ S$\\  \hline
$P_{9}$&$0.042  \ i \ S $&$0.044 \ i  \ U$&$0.101 \ S$&$ \ \ \  $& $0.0044 \ i  \ S$&$0.0043 \ i \ U$&$0.0101 \ S$\\  \hline
$P_{10}$&$0.041  \ i  \ S$&$0.043  \ i \ U $&$0.100 \ S$&$ \ \ \  $&  $0.0043  \ i  \ S$&$0.0042 \ i  \ U$&$0.01 \ S$\\  \hline
$P_{11}$&$0.033 \ S  $&$0.036 \ S$&$0.989 \ S$&$ \ \ \ $&  $0.0042 \ i  \ S$& $0.0036 \ S$&$0.099 \ S$\\  \hline
$P_{12}$&$0.032 \ S  $&$0.035 \ S$ &$0.096 \ S$&$ \ \ \  $&  $0.0041 \ i  \ S$&$0.0035 \ S$ &$0.0096 \ S$\\  \hline
 $P_{13}$&$0.030  \ i \ S $&$0.031 \ i \ U$ &$0.085 \ S$&$ \ \ \  $&$0.003 \ S$&$0.0031 \ i \ U$ &$0.0086 \ S $\\  \hline
$P_{14}$ & $0.027  \ i \ S $&$0.029  \ i  \ U$&$0.081 \ S $&$ \ \ \  $&$0.0026 \ S$&$0.003 \ i  \ U$&$0.0081 \ S $\\  \hline
$P_{15}$&$0.025 \ S $&$0.027 \ S $&$0.077 \ S  $&$ \ \ \  $&$0.0025 \ S $&$0.0027  \ S  $&$0.0077 \ S $\\  \hline
$P_{16}$&$0.024 \ S $&$0.026 \ S $&$0.076 \ S  $&$ \ \ \  $&$0.002 \ S $&$0.003  \ S  $&$0.0076 \ S $\\  \hline
$P_{17}$&$0.02 \ S $&$0.02 \ S $&$0.063 \ S  $&$ \ \ \  $&$0.0024 \ S $&$0.0021  \ S  $&$0.0063 \ S $\\  \hline
$P_{18}$&$0.018 \ S $&$0.02 \ S $&$0.059 \ S  $&$ \ \ \    $&$0.0024 \ S $&$0.0025  \ S  $&$0.006 \ S $\\  \hline
\end{tabular}
\caption{\label{tab:table1} Eighteen first periods associated to stable (S) and unstable (U) eigenvalues (minutes) for A) Left panel: $B_{0}=10G$ with A1) left column: weak helicity, A2) middle column: moderate helicity, A3) right column: strong helicity and B) Right panel: $B_{0}=100G$ with B1, B2, B3 the same as in A. Larger order modes were discarded.}
\end{table*}

\begin{table*}
\begin{tabular}{ccccccc}
\hline
$P_{i}$&$weak $&$moderate$&$strong  \ \ \ \ $&$weak $&$moderate$&$strong$\\  \hline
$P_{1}$&$\xi_{\parallel} \gg \xi_{\perp} \mapsto 0  \ S; \ IP $&$\xi_{\parallel}  > \xi_{\perp}  \ S; \ IP  $&$\xi_{\parallel}  \geq \xi_{\perp}
  \ P \ \ \ \ $&$ \xi_{z}\gg  \xi_{\phi}   \sim   \xi_{r}\mapsto 0  $&$ \xi_{z}\gg  \xi_{\phi}   \sim   \xi_{r} $&$  \xi_{r} \leq \xi_{\phi}; \xi_{z}  \mapsto 0  $\\  \hline
$P_{2}$&$\xi_{\parallel} \gg \xi_{\perp} \mapsto 0  \ S; \ IP $&$\xi_{\parallel}  > \xi_{\perp}  \ S; \ IP  $&$\xi_{\parallel}  \geq \xi_{\perp}  \ P \ \ \ \ $&$ \xi_{z}\gg  \xi_{\phi}   \sim   \xi_{r}\mapsto 0  $&$ \xi_{z}\gg \xi_{\phi}   \sim   \xi_{r}  $&$   \xi_{r} \leq \xi_{\phi}; \xi_{z}  \mapsto 0  $\\  \hline
$P_{3}$&$\xi_{\parallel} \gg \xi_{\perp} \mapsto 0  \ S; \ IP $&$\xi_{\parallel}  > \xi_{\perp}  \ S; \ IP  $&$\xi_{\parallel}  \geq \xi_{\perp}  \ P \ \ \ \ $&$ \xi_{z}\gg  \xi_{\phi}   \sim   \xi_{r}\mapsto 0  $&$ \xi_{z}\gg  \xi_{\phi}   \sim   \xi_{r} $&$  \xi_{r} \leq \xi_{\phi}; \xi_{z}  \mapsto 0 \ $\\  \hline
$P_{4}$&$\xi_{\parallel} \gg \xi_{\perp}  \mapsto 0 \ S; \ IP $&$\xi_{\parallel}  > \xi_{\perp}  \ S; \ IP  $&$\xi_{\parallel}  \geq \xi_{\perp}  \ P \ \ \ \ $&$ \xi_{z}\gg  \xi_{\phi}   \sim   \xi_{r}\mapsto 0  $&$ \xi_{z}\gg  \xi_{\phi}   \sim   \xi_{r}  $&$  \xi_{r} \leq \xi_{\phi}; \xi_{z}  \mapsto 0 $\\  \hline
$P_{5}$&$\xi_{\parallel} \gg \xi_{\perp} \mapsto 0 \ S;  \ IP $&$\xi_{\parallel}  > \xi_{\perp}  \ S; \ IP  $&$\xi_{\parallel}  \geq \xi_{\perp}  \ P  \ \ \ \ $&$ \xi_{z}\gg  \xi_{\phi}   \sim   \xi_{r}\mapsto 0  $&$ \xi_{z}\gg  \xi_{\phi}   \sim   \xi_{r}  $&$ \xi_{r} \leq \xi_{\phi}; \xi_{z}  \mapsto 0  $\\  \hline
$P_{6}$&$\xi_{\parallel} \gg \xi_{\perp}  \mapsto 0 \ S; \ IP $&$\xi_{\parallel}  > \xi_{\perp}   \ S; \ IP  $&$\xi_{\parallel}  \geq \xi_{\perp}  \ P \ \ \ \ $&$ \xi_{z}\gg  \xi_{\phi}   \sim   \xi_{r}\mapsto 0  $&$ \xi_{z}\gg  \xi_{\phi}   \sim   \xi_{r} $&$ \xi_{r} \leq \xi_{\phi}; \xi_{z}  \mapsto 0  $\\  \hline \hline
$P_{7}$&$\xi_{\perp} \gg \xi_{\parallel} \mapsto 0  \ F; \ P $&$\xi_{\perp} > \xi_{\parallel}  \ F;  \ P$ &$\xi_{\perp}  \geq \xi_{\parallel}  \ IP \ \  \ \ $&$\xi_{r}   \sim   \xi_{\phi}\gg   \xi_{z}\mapsto 0$&$\xi_{r}   \sim  \xi_{\phi}\gg \xi_{z} \mapsto 0  $&$\xi_{z} > \xi_{r}  > \xi_{\phi}  $\\  \hline
$P_{8}$&$\xi_{\perp} \gg \xi_{\parallel}\mapsto 0   \ F; \ P $&$\xi_{\perp} > \xi_{\parallel}   \ F; \ P$&$\xi_{\perp}  \geq \xi_{\parallel}  \ IP \ \ \ \ $&$\xi_{r}   \sim   \xi_{\phi}\gg   \xi_{z}\mapsto 0 $&$\xi_{r}   \sim   \xi_{\phi}\gg  \xi_{z}\mapsto 0  $&$\xi_{z} > \xi_{r}> \xi_{\phi}  $ \\  \hline
$P_{9}$&$\xi_{\perp} \gg \xi_{\parallel}\mapsto 0  \ F;  \ P $&$\xi_{\perp} > \xi_{\parallel}   \ F; \ P$&$\xi_{\perp}  \geq \xi_{\parallel}  \ IP \ \ \ \ $&$\xi_{r}  \sim   \xi_{\phi}\gg  \xi_{z}\mapsto 0  $&$\xi_{r}  \sim  \xi_{\phi}\gg  \xi_{z}\mapsto 0  $&$\xi_{z} > \xi_{r} > \xi_{\phi}  $\\  \hline
$P_{10}$&$\xi_{\perp} \gg \xi_{\parallel}\mapsto 0 \ F;   \ P $&$\xi_{\perp} > \xi_{\parallel}  \ F;  \ P$&$\xi_{\perp}  \geq \xi_{\parallel}  \ IP \ \ \ \ $& $\xi_{r}   \sim   \xi_{\phi}\gg  \xi_{z}\mapsto 0  $&$\xi_{r}   \sim   \xi_{\phi}\gg  \xi_{z}\mapsto 0  $&$\xi_{z}> \xi_{r} > \xi_{\phi}  $\\  \hline
$P_{11}$&$\xi_{\perp} \gg \xi_{\parallel}\mapsto 0  \ F;  \ P $&$\xi_{\perp} > \xi_{\parallel}  \ F;  \ P$&$\xi_{\perp}  \geq \xi_{\parallel}  \ IP \ \ \ \ \ $&$\xi_{r}   \sim   \xi_{\phi}\gg   \xi_{z}\mapsto 0  $&$\xi_{r}   \sim   \xi_{\phi} \gg  \xi_{z}\mapsto 0  $&$\xi_{z} > \xi_{r} > \xi_{\phi}  $ \\  \hline
$P_{12}$&$\xi_{\perp} \gg \xi_{\parallel}\mapsto 0  \ F;  \ P $&$\xi_{\perp} > \xi_{\parallel}   \ F; \ P$&$\xi_{\parallel}  \geq \xi_{\perp}  \ IP \ \ \ \ $&$\xi_{r}   \sim  \xi_{\phi}\gg   \xi_{z}\mapsto 0  $&$\xi_{r} \sim  \xi_{\phi}\gg \xi_{z}\mapsto 0  $ &$  \xi_{r}>  \xi_{\phi} > \xi_{z}   $ \\  \hline
 $P_{13}$&$\xi_{\perp} \gg \xi_{\parallel} \mapsto 0  \ F; \ P $&$\xi_{\perp} > \xi_{\parallel}   \ F; \ P$&$\xi_{\perp}  \geq \xi_{\parallel}  \ IP \ \ \ \ $&$\xi_{r}\sim    \xi_{\phi}\gg  \xi_{z}\mapsto 0  $&$\xi_{r}   \sim  \xi_{\phi}\gg  \xi_{z}\mapsto 0 $&$\xi_{z} > \xi_{r} > \xi_{\phi}  $\\  \hline
$P_{14}$&$\xi_{\perp} \gg \xi_{\parallel}\mapsto 0   \ F; \ P $&$\xi_{\perp} > \xi_{\parallel}  \ F;  \ P$ &$\xi_{\perp}  \geq \xi_{\parallel}  \ IP \ \ \ \ $&$\xi_{r}  \sim   \xi_{\phi}\gg   \xi_{z}\mapsto 0 $&$\xi_{r}  \sim  \xi_{\phi}\gg  \xi_{z}\mapsto 0 $&$\xi_{z} > \xi_{r} > \xi_{\phi}  $\\  \hline
$P_{15}$&$\xi_{\perp} \gg \xi_{\parallel} \mapsto 0 \ F;  \ P $&$\xi_{\perp} > \xi_{\parallel}  \ F;  \ P$&$\xi_{\parallel}  \geq \xi_{\perp}  \ P \ \ \ \ $&$\xi_{r}  \sim   \xi_{\phi}\gg \xi_{z}\mapsto 0 $&$\xi_{r}  \sim   \xi_{\phi}\gg \xi_{z}\mapsto 0$&$  \xi_{r}>  \xi_{\phi} > \xi_{z}   $ \\  \hline
$P_{16}$&$\xi_{\perp} \gg \xi_{\parallel} \mapsto 0 \ F;  \ P $&$\xi_{\perp} > \xi_{\parallel}  \ F;  \ P$&$\xi_{\parallel}  \geq \xi_{\perp}  \ P \ \ \ \ $&$\xi_{r}  \sim   \xi_{\phi}\gg \xi_{z}\mapsto 0  $&$\xi_{r} \sim    \xi_{\phi}\gg  \xi_{z}\mapsto 0  $&$  \xi_{r}>  \xi_{\phi} > \xi_{z}   $ \\  \hline
$P_{17}$&$\xi_{\perp} \gg \xi_{\parallel} \mapsto 0  \ F; \ P $&$\xi_{\perp} > \xi_{\parallel}  \ F;  \ P$&$\xi_{\parallel}  \geq \xi_{\perp}  \ P \ \ \ \ $&$\xi_{r} \sim  \xi_{\phi}\gg    \xi_{z}\mapsto 0$&$\xi_{r}   \sim  \xi_{\phi}\gg  \xi_{z}\mapsto 0 $&$  \xi_{r}>  \xi_{\phi} > \xi_{z}   $ \\  \hline
$P_{18}$&$\xi_{\perp} \gg \xi_{\parallel} \mapsto 0  \ F; \ P $&$\xi_{\perp} > \xi_{\parallel}   \ F; \ P$ &$\xi_{\parallel}  \geq \xi_{\perp}  \ P \ \ \ \ $&$\xi_{r}  \sim   \xi_{\phi}\gg  \xi_{z}\mapsto 0  $&$\xi_{r}  \sim  \xi_{\phi}\gg  \xi_{z}\mapsto 0 $&$  \xi_{r}>  \xi_{\phi} > \xi_{z}   $\\  \hline
\end{tabular}
\caption{\label{tab:table2} First Panel: Intensity relationship between the tangential and normal to the field components of the eighteen   first periods for $B_{0}=10G$ and for  weak (first column), moderate (second column) and strong helicity (third column) cases.  The (P) indicates in phase and (IP) indicates inverted phase. Second Panel: Intensity relationship between the cylindrical  components of the eighteen   first periods for $B_{0}=10G$ and for  weak (first column), moderate (second column) and strong helicity (third column) cases.}
\end{table*}

\begin{table*}
\begin{tabular}{cccccccc}
\hline
$R$&$L $&$R/2L$&$Twist=bR \ \ \ \ \ \ \ \ \ $&$R$&$L $&$R/2L$&$Twist=bR$ \\  \hline
$0.01$& $ 9.05 \ 10^{7} $&$0.005 $&$  0.028\ \ \ \ \ \ \ \ \ $&$0.01$& $ 8.07 \ 10^{7} $&$0.005 $&$  0.28$\\  \hline
$0.02$&$ 9.04 \ 10^{7}  $&$ 0.01  $&$ 0.057  \ \ \ \ \ \ \ \ \ $&$0.015$&$ 8.32 \ 10^{7}  $&$ 0.008  $&$ 0.43  $\\  \hline
$0.03$&$ 9.02 \ 10^{7} $&$ 0.015 $&$ 0.085 \ \ \ \ \ \ \ \ \ $&$0.02$&$ 7.86 \ 10^{7} $&$ 0.011 $&$ 0.57$\\  \hline
$0.04  $&$ 8.99 \ 10^{7}  $&$ 0.02 $&$ 0.11\ \ \ \ \ \ \ \ \ $&$0.025  $&$ 7.38 \ 10^{7}  $&$ 0.015 $&$ 0.71$\\  \hline
$0.05  $&$ 8.96 \ 10^{7}  $&$ 0.025 $&$ 0.14\ \ \ \ \ \ \ \ \ $&$0.03  $&$ 6.88 \ 10^{7}  $&$ 0.02 $&$ 0.85$\\  \hline
$0.06  $&$ 8.9 \ 10^{7}  $&$ 0.03 $&$ 0.17\ \ \ \ \ \ \ \ \ $&$0.04  $&$ 5.97 \ 10^{7}  $&$ 0.03 $&$ 1.13$\\  \hline
$0.1  $&$ 8.7 \ 10^{7}  $&$ 0.05 $&$ 0.28\ \ \ \ \ \ \ \ \ $&$0.05  $&$ 5.2 \ 10^{7}  $&$ 0.04 $&$ 1.42$\\  \hline
 \hline
\end{tabular}
\caption{\label{tab:table5} First Panel - Stable case: Variation of the Radius with the Twist for weak helicity $b=0.05$ and  $B_{0}=10G$. Second Panel - Unstable case: Variation of the Radius with the Twist for moderate helicity $b=0.5$ and  $B_{0}=10G$.}
\end{table*}

\section{Appendix: Generalized potential energy terms}

From the procedure described above and extensively exemplified in Paper I we 
can  obtain -laboriously but in a straightforward way- the  explicit terms for the energy principle given in eq.~\ref{10}:
$$\delta^{2} W_{p}= \delta^{2} W_{c}+\delta^{2} W_{m}+\delta^{2} W_{hc}+\delta^{2} W_{r} $$
where the right side of the equation corresponds to the compressional, magnetic, heat conduction and radiative contributions respectively.
The compressional term $ \delta^{2} W_{c}=\delta^{2} W_{c1}+\delta^{2} W_{c2}$ has an additional contribution ($\delta^{2} W_{c2}$) with respect to Bernsteins principle:
$$ \delta^{2} W_{B}= \delta^{2} W_{c1}+ \delta^{2} W_{m}$$

 $$
 \delta^{2} W_{c1}=\frac{1}{2}\int_{-1}^{1}dz \beta\left\{ T_{0} \rho_{0} (1-m) \frac{\xi_{r}^{2}}{R^{2}}-\frac{m}{R}T_{0}\left(\Delta\frac{d\rho_{0}}{ds}\xi_{z}\xi_{\phi}+\rho_{0}\left(\frac{\xi_{r}\xi_{\phi}}{R} -\right.\right.\right.$$

$$ \left. \left. -\frac{m}{R}\xi_{\phi}^{2}+\frac{d\xi_{\phi}}{dz}\right)\right)+\Delta  \frac{dT_{0}}{ds}\left(\Delta  \frac{d\rho_{0}}{ds}\xi_{z}^{2}+\rho_{0}(\frac{\xi_{r}\xi_{\phi}}{R} -\frac{m}{R}\xi_{\phi}\xi_{z}+\xi_{z}\frac{d\xi_{z}}{dz}) \right) +
$$
$$
+T_{0}
\left(\Delta^{2}\frac{d^{2}\rho_{0}}{ds^{2}}\xi_{z}^{2}+\Delta\frac{d\rho_{0}}{ds}
\xi_{z}\frac{d\xi_{z}}{dz}+\rho_{0}(\frac{\xi_{z}}{R}\frac{d\xi_{r}}{dz}-
\frac{m}{R}\xi_{z}\frac{d\xi_{\phi}}{dz} 
+\xi_{z}\frac{d^{2}\xi_{z}}{dz^{2}})+\right.
$$
$$\left. \Delta\frac{d\rho_{0}}{ds}(\frac{\xi_{r}\xi_{\phi}}{R} -\frac{m}{R}
\xi_{\phi}\xi_{z}+\xi_{z}\frac{d\xi_{z}}{dz}) \right) 
$$
The magnetic contribution is:
$$ \delta^{2} W_{m}=C_{1}\left\{ \beta \Delta \left (\frac{m}{R}B_{\phi}B_{z}\xi_{r}
\xi_{\phi}-B_{\phi}B_{z}\frac{d\xi_{r}}{dz}\xi_{z}\right. \right. $$
$$
-\left.  (\frac{B_{\phi}B_{z}}{R}+B_{\phi}\frac{dB_{z}}{dr})
 \xi_{r}^{2}+(\frac{m}{R}B_{\phi}B_{z}\xi_{r}\xi_{\phi}+B_{\phi}
 \frac{dB_{\phi}}{dr} \xi_{r}^{2})\right) -$$$$
-\beta \left( (B_{z}^{2}\frac{d\xi_{r}^{2}}{dz}+\frac{m^{2}}{R^{2}}
B_{\phi}^{2}\xi_{r}^{2})+(B_{\phi}^{2}\frac{d\xi_{z}^{2}}{dz}+2B_{\phi}+
\frac{dB_{\phi}}{dr}\xi_{r}\frac{d^{2}\xi_{z}}{dz}+
 \frac{dB_{\phi}^{2}}{dr}\xi_{r}^{2}+B_{z}^{2}\frac{d^{2}
\xi_{\phi}}{dz})+\right. 
$$$$
\left. \left. \left ( (B_{z}+R\frac{dB_{z}}{dr})^{2}\frac{\xi_{r}^{2}}{R^{2}}-
 2\frac{m}{R^{2}}B_{z}(B_{z}+R\frac{dB_{z}}{dr})
\xi_{r}\xi_{\phi}+(\frac{m}{R}B_{z})^{2}\xi_{\phi}^{2}+
(\frac{m}{R}B_{\phi})^{2}
\xi_{z}^{2}\right) \right) \right\} 
$$
The heat conduction term results:
$$\delta^{2} W_{hc}=-C_{2}\left\{5\frac{T_{0}^{3/2}}{\Delta}
\frac{dT_{0}}{ds}T_{1}\frac{dT_{1}}{dz}+T_{1}^{2}
\left( -T_{0}^{5/2}(\frac{mb}{\Delta})^{2}\frac{15}{4}T_{0}^{1/2}
\frac{dT_{0}^{2}}{ds}+\right.\right.  
$$
$$
\left. \left. +\frac{5}{2}T_{0}^{3/2}\frac{d^{2}T_{0}}{ds^{2}}\right)
+\frac{1}{\Delta^{2}}T_{0}^{5/2}T_{1}\frac{d^{2}T_{1}}{dz^{2}}\right\}
$$
The new compressional contribution is expressed as:
$$\delta^{2} W_{c2}=-\beta \left(\frac{m}{R} \rho_{0}\xi_{\phi}T_{1}+\Delta\frac{d \rho_{0}}{ds}
\xi_{z}T_{1}+\rho_{0}\xi_{z}\frac{dT_{1}}{dz}\right)
$$
and the term associated to radiation results:
$$
\delta^{2} W_{r}=-\alpha T_{1}^{2}\rho_{0}^{2}T_{0}^{\alpha-1}-\beta \left(\frac{m}{R} \rho_{0}\xi_{\phi}T_{1}+\Delta\frac{d \rho_{0}}{ds}
\xi_{z}T_{1}+\rho_{0}T_{1}\frac{d\xi_{z}}{dz}+
\frac{\rho_{0}}{R} \xi_{r}T_{1}\right)$$

\noindent
where the following changes were made:
$$\rho\rightarrow\frac{\rho}{\rho_{t}};  \   \
T\rightarrow\frac{T}{T_{t}}; \   \
B_{\phi,z}\rightarrow\frac{B_{\phi,z}}{B_{0}}; \   \ b\rightarrow b S$$
$$ r,z\rightarrow\frac{r,z}{S}; \  \  \delta^{2}
W_{p}\rightarrow\  \delta^{2} W_{p}/\left(\chi
T_{t}^{\alpha+1}\rho_{t}^{2}L/m_{p}^{2}\right),$$
$S=\Delta L$ and the   non--dimensional constants: $$C_{1}=
\rho_{t}^{2}T_{t}^{\alpha+1}B_{0}^{2}/(\mu_{0}k_{B} T_{t}n_{e}); \  \
C_{2}=c  T_{t}^{\frac{7}{2}-\alpha}/(S^{2} n_{e}^{2}).$$
were used. All the  quantities were defined as  in Paper I.

 \label{lastpage}

\end{document}